# Contextuality as the Key to understand Quantum Paradoxes

Marian Kupczynski

Département de l'Informatique, UQO, Case postale 1250, succursale Hull, Gatineau. QC, Canada J8X 3X7

**\* Correspondence:**
marian.kupczynski@uqo.ca

Abstract: This short note is an introduction to the slides from our talks given in 2018 during the Advanced School of Quantum Foundations and Quantum Computation in Joao Pessoa , Brazil. In these talks, addressed to participants with a limited knowledge of quantum foundations, we defined locality, causality, randomness, counterfactual definiteness and we explained EPR-Bohm paradoxes. In particular we discussed in great detail various Bell-type inequalities and the implications of their violation in spin polarisation correlation experiments. Finally we explained why and how the predictive completeness of quantum mechanics may be tested by more careful examination of the time-series of experimental data.

Quantum mechanics (QM) has led to spectacular technological developments, including the discovery of new constituents of matter and new materials: however, there is still no consensus regarding its interpretation and limitations [1-2].

The incorrect interpretations of QM and incorrect mental models of invisible details of quantum phenomena lead to paradoxes and speculations about quantum weirdness and quantum magic. Since we are witnessing now a second quantum technological revolution it is important to realize that without a correct understanding of quantum foundations no real technological progress is possible.

In our talks we advocated a contextual statistical interpretation [3-12] of QM which is free of paradoxes. According to this interpretation a quantum state is not an attribute of an individual physical system which may be changed instantaneously. Quantum state/ wave function is a mathematical entity representing an equivalence class of subsequent preparations of the physical systems. Quantum states together with self-adjoint operators representing physical observables are used to make probabilistic predictions for a statistical scatter of measured values of these observables.

A probability is a subtle notion and may have a different meaning for different scientists and philosophers [10, 13-15]. For us a probability is the objective property of random experiments in which empirical frequencies stabilize. It is neither a property of a physical system nor a degree of belief of conscious agents. Therefore a probabilistic description of various phenomena can hardly be considered as a complete description of individual physical systems [2].

This is why Einstein believed that a more detailed description of individual physical systems should be found in which quantum probabilities could be deduced [3-4]. Bohr insisted that quantum probabilities are irreducible and that QM provides a complete description of individual physical systems [16-19].

In 1927 Heisenberg [20] demonstrated the uncertainty principle according to which one may not measure simultaneously, with arbitrary accuracy, a linear momentum *p* and a position x of a sub-atomic particle : $\Delta x \Delta p \leq h$ where *h* is a Planck constant. Heisenberg did not specify the exact meaning of the uncertainties $\Delta x \Delta p$. The uncertainty principle is valid with some modifications for any pair of complementary quantum observables described by non- commuting self-adjoint operators [21].

The more precise formulation of the uncertainty principle was given by Kennard [22]. Let us consider two experiments using two identically prepared beams of particles. In one experiment we measure their linear momenta and in another the positions. The measured values in both experiments have a statistical scatter and experimental data may described by specific probability distributions. For any initial preparation of particles, described by any quantum state , standard deviations of these probability distributions have to obey the inequality : $\sigma_x \sigma_p \leq \hbar/2$ where $\hbar = h/2\pi$. In this interpretation we are talking about the statistical scatter of measurement outcomes and not what is a position and a linear momentum of an electron if no measurements are done..

By no means Heisenberg uncertainty principle can lead to the conclusion that an electron can be <u>at the same time here and a meter away</u> what is incorrectly claimed by some authors.

According to Copenhagen interpretation of QM all the speculations about sharp unmeasured values of linear momenta and positions of sub-atomic particles are meaningless. For Einstein QM does not provide a complete description of individual physical systems and a mental image of a point-like electron , being somewhere, and having a sharp value of a linear momentum at a given moment of time is not in conflict with the Heisenberg uncertainty principle and with purely statistical interpretation of QM [3-4].

In 1935 Einstein, Podolsky and Rosen [23] analyzed two physical systems I+II, in distant locations, having strongly correlated properties after they interacted in the past. If one decided to measure a linear momentum or a position of the system I , then using the outcomes of these measurements one might evaluate the linear momentum or the position of the distant system II without disturbing it in any way. This is called the EPR paradox. This paradox was rephrased by Bohm [24] in terms of the measurements of the electron's spin polarization in different directions. A detailed discussion of EPR and EPR-B paradoxes may be found in [7] and in the attached slides.

QM predicts strong correlations between spin projections of entangled pairs of electrons and photons measured in distant laboratories by Alice in Bob in ideal EPRB experiment.

Distant outcomes should be perfectly correlated or anti-correlated in different experimental settings and at the same time they are believed to be produced in an irreducible random way. Such two requirements are impossible to satisfy at the same time [25-28].

Bell [29-30] abandoned irreducible randomness and proposed Local Realistic Hidden Variable Model ( LRHVM) in which the outcomes are predetermined at the source. Clauser and Horne [31] abandoned predetermination and proposed Stochastic Hidden Variable Model ( SHVM). In these models , correlations between distant outcomes coded ±1 have to obey Bell-CHSH inequalities [32] which are for some settings violated by quantum predictions and by the experimental data.

Bell correctly pointed out that his model [29] may not reproduce quantum predictions for the ideal EPRB experiment and believed that no other locally causal explanation of quantum correlations was possible. This is why the violation of various inequalities in real spin polarization correlation experiments (SPCE) is the source of various speculations about a mysterious *quantum nonlocality* or about *super-determinism* limiting experimentalists' freedom to choose their experimental settings.

Fortunately an ideal EPRB experiment may not be performed and imperfect correlations between Alice's and Bob's clicks in real SPCE may be explained in a causally local way without evoking the quantum magic [2 ,8, 27, 28 , 33-34].

We are not going to repeat in this introduction detailed arguments against *quantum nonlocality* and magical explanations of correlations. We limit ourselves to few conclusions and more details may be found in several references [49-64] and in the slides which follow.

1. In SPCE for different pairs of the settings we have 4 incompatible random experiments. LRHVM and SPHVM are using a unique probability space and a joint probability distribution to describe all these experiments what is only possible in rare circumstances [8,14-15,33-37,41-42,57-58].
2. In QM interactions of instruments with the physical systems during the measurement process may not be neglected and the outcomes are not passively registered pre-existing values of the physical observables. LRHVM and SHVM suffer therefore from theoretical "*contextuality loophole*" [52-54] since they fail to include correctly setting dependent variables describing the measuring instruments at the moment of the measurement.
3. Bell and CHSH inequalities are trivial algebraic properties of experimental spreadsheets [28,38-40,42,49,56,60,62] containing quadruplets of ±1 which are in fact samples drawn from a statistical population described by some joint probability distribution of 4 compatible random variables. Such probability distribution does not exist in SPCE.
4. To test predictable completeness of QM time series of experimental data have to be studied more in detail than it is presently done [46,65-68]

Therefore the violation of various Bell-type inequalities is not surprising and does not justify the speculations about the existence of spooky influences in Nature. The impossibility of treating a quantum state as an attribute of an individual quantum system should be well understood by quantum information community.

We finish this introduction by an ordered list of topics which are discussed in the attached slides.

- Introduction. Bertrand Paradox. Probability as a contextual property of a random experiment.
- Copenhagen interpretation of QM. Statistical contextual interpretation of QM. EPR-Bohm paradoxes explained.
- Locality, causality, randomness, counterfactual definiteness (CFD) and physical reality. Local realism versus contextuality,
- Probabilistic models, experimental protocols, expectation values, joint probability distributions and correlations, distant correlations, common cause and generalised joint probability distributions.
- Spin polarization correlation experiments. Quantum nonlocal correlations, local realistic hidden variable models, stochastic hidden variable models, Bell, CHSH and CH inequalities
- Detailed discussion of LRHVM and SHVM and of impossible experimental protocols they imply.
- Contextuality and locally causal description of quantum correlations and experimentalists' *freedom of choice.*
- Conclusions and lessons for the quantum information community.
- How to test predictable completeness of QM. Purity tests . Search for fine structures in time series of experimental data which might have been be averaged out during standard statistical analysis.
- References.

Our list of references contains only the papers which give more detailed discussion of the topics presented in this introduction and in the slides. Much more complete lists of the references giving a due credit to other authors may be found in the papers listed below.

# III Adanced School of Quantum Foundations and Quantum Computation
## Joao Pessoa. Brazil, 26/11-07/12 2018

# Contextuality as the key to understand quantum paradoxes

Marian Kupczynski
UQO
marian.kupczynski@uqo.ca

http://w4.uqo.ca/kupcma01/homepage.htm

Marian Kupczynski

---

## Some my older papers

- M. Kupczynski , Pitovsky model and complementarity, Phys.Lett. A **121**, 51(1987)
- M.Kupczynski, Seventy years of the EPR paradox AIP Conf. Proc. 861,516 (2006)
- M. Kupczynski, Entanglement and Bell inequalities J. Russ. Laser Res. **26**, 514(2005)
- M. Kupczynski , EPR paradox ,locality and completeness of quantum, AIP Conf. Proc. **962** ,274 (2007).
- M.Kupczynski , Entanglement and quantum nonlocality demystified AIP Conf. Proc. vol 1508 253 (arXiv : 0710.3550 [quant-ph])
- M.Kupczynski, Causality and local determinism versus quantum nonlocality, J. Phys.: Conf. Ser. 504 012015 doi:10.1088/1742-6596/504/1/012015
- M. Kupczynski, Bell Inequalities, Experimental Protocols and Contextuality, Found. Phys. (12 Dec 2014), doi:10.1007/s10701-014-9863-(2014) *(arXiv:1411.7085)*
- M.Kupczynski, EPR Paradox, Quantum Nonlocality and Physical Reality , J. Phys.: Conf. Ser. 701 012021 (2016)

Marian Kupczynski

---

## Main references

- Kupczynski M., Bertrand's paradox and Bell's inequalities, Phys.Lett. A **121,** 205 (1987).
- Kupczynski, M.,Can we close the Bohr–Einstein quantum debate? Phil. Trans. R. Soc. A . 375: 0160392.http://dx.doi.org/10.1098/rsta.2016.0392 (arXiv:1602.02959)
- Kupczynski M,Quantum mechanics and modelling of physical reality. Phys. Scr. 93 (2018) 123001 (10pp) https://doi.org/10.1088/1402-4896/aae212 ( arXiv:1804.02288 )
- Kupczynski,M. Closing the door on quantum nonlocality. Entropy 2018, 20, 877.DOI: 10.3390/e20110877 (Open Access)
- Khrennikov, A. Contextual Approach to Quantum Formalism; Springer: Dordrecht, Netherlands,2009

Marian Kupczynski

---

## Tentative topics

- Introduction. Bertrand Paradox. Probability as a contextual property of a random experiment.
- Copenhagen interpretation of QM. Statistical contextual interpretation of QM. EPR-Bohm paradoxes explained.
- Locality, causality, randomness, counterfactual definiteness (CFD) and physical reality. Local realism versus contextuality,
- Probabilistic models, experimental protocols, expectation values, joint probability distributions and correlations, distant correlations, common cause and generalised joint probability distributions.
- Spin polarization correlation experiments (SPCE). Quantum nonlocal correlations, local realistic hidden variable models (LRHV), stochastic hidden variable models (SHVM), Bell, CHSH and CH inequalities
- Detailed discussion of LRHVM and SHVM and of impossible experimental protocols they imply. Contextuality and locally causal description of quantum correlations and experimentalists' *freedom of choice*. Conclusions and lessons for the quantum information community.

Marian Kupczynski

---

## Other references

Each of the papers listed above contains its own list of references. I asked also Andrei Khrennikov and Hans de Raedt to recommend some of their papers which can be easily accessed by the students. Their recommendations are included on two last slides of these series of talks.

A theoretical loophole in various proofs of Bell inequalities which I will explain in my talks was called by Theo Nieuwenhuizen : CONTEXTUALITY LOOPHOLE.

- T.M. Nieuwenhuizen , Where Bell went wrong AIP Conf. Proc. 1101, 127 (2009)
- T.M. Nieuwenhuizen , Is the contextuality loophole fatal for the derivation of Bell inequalities Found. Phys. 41 , 580 (2011)

Marian Kupczynski

---

## FEW WORDS OF INTRODUCTION

- In spite of the fact that statistical predictions of quantum theory (QT) can only be tested, if the large amount of data is available the claim has been made that the QT provides the most complete description of the individual physical system.
- Einstein's opposition to this claim and the paradox he presented in the article written together with Podolsky and Rosen in 1935 inspired generations of physicists in their quest for better understanding of QT.
- Without deep understanding of the character and limitations of QT one may not hope to find a meaningful unified theory of all physical interactions, manipulate qubits or construct a quantum computer.

Marian Kupczynski

## Points of view

- "There is no quantum world . There is only an abstract quantum mechanical description". N.Bohr
- "The physical content of quantum mechanics is exhausted by its power to formulate statistical laws governing observations under the conditions specified in plain language" . N. Bohr
- "I am therefore inclined to believe that the description of quantum mechanics has to be regarded as an incomplete and indirect description of reality to be replaced by some later date by a more complete and direct one." A. Einstein

Marian Kupczynski

## My philosophical stand point

- The moon does continue to exist even, if we do not look at it.
- Probability is an objective contextual property of some random experiments and not a subjective belief of some intelligent agents.
- There is no need to reject causality and locality in order to explain EPR correlations and the violation of Bell-type inequalities
- In order to verify the completeness of QT more detailed analysis of experimental time-series is needed

Marian Kupczynski

## Quantum Magic

- "We roll a pair of dice. Each die on its own is random and fair, but its entangled partner somehow always gives the correct matching outcome. Such behavior has been demonstrated and intensively studied with real entangled particles" PRL 2001
- Quantum nonlocality , whereby particles appear to influence one another instantaneously even though they are widely separated …..is a well established experimental fact. PRL 2011 → MYSTERY
- Correlations are coming out of space time! Science 2013
- A particle passes through two neighboring slits at the same time. Therefore, an electron, is indeed both here and a meter to the right of here. 2014 (IMPOSSIBLE!!!)

Marian Kupczynski

## What does it mean to understand

- In everyday life we are asking many questions :"Why..." and we are getting or giving answers: "Because..." but to any answer "Because..." there is immediately another question: "Why..." and so on so on.
- Therefore in physics and in any other brunch of science we are only able to detect, explain and predict regularities in the phenomena perceived by us in the surrounding world or created by us in the laboratory.
- We believe that there should be some universal laws and mathematical models explaining all these phenomena.
- To explain the properties of visible objects and phenomena we have to introduce some more elementary invisible objects and sub phenomena .

Marian Kupczynski

## Conclusion

- R.Feynman :
    "Nobody understands quantum mechanics"

- IF WE CONTINUE TO USE IMPRECISE LANGUAGE AND INCORRECT IMAGES AND ANALOGIES NOBODY WIL EVER DO (MK).

Marian Kupczynski

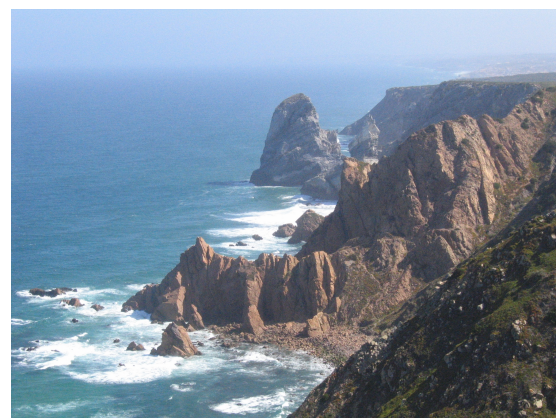

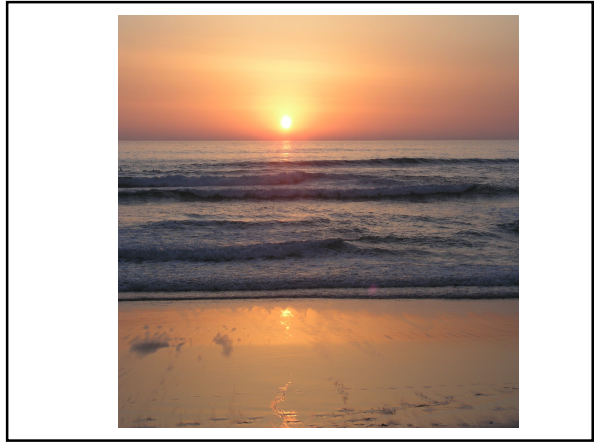

## Sub Phenomena 1

A ball 1 is entering a black box and "identical ball" is out

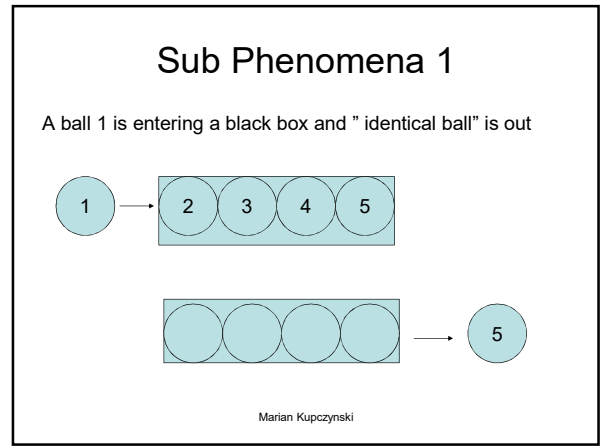

Marian Kupczynski

## View From Cabo da Roca

1) Solid rocks having constant attributes: such as a shape, distances between different points on the surface
2) Water characterized by its salinity (attribute) and by its average temperature
3) Reflected light on the rocks or the water characterized by its intensity or color.
4) Waves on the water defined by their amplitude, length and changeable velocity of the wave front
5) Interference of waves, refraction and diffraction of light
6) Turbulences on the water, clouds, lightings etc
7) Planets and stars

Marian Kupczynski

## Sub Phenomena 2 or 3

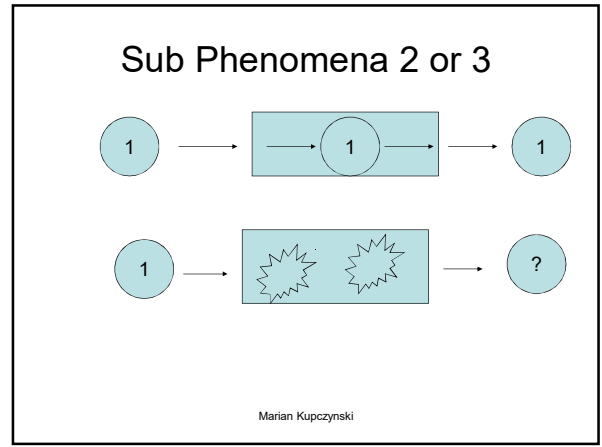

Marian Kupczynski

## Phenomena

1. Corpuscular (Throwing a stone from Cabo du Roca)
- Well defined individuality and localization of the object
- Continuous trajectory. Motion described by second order ordinary differential equations
- Linear momentum and kinetic energy carried by object
2. Wave (Waves hitting the shore)
- Considerable space extension
- Motion= deformations of the surface in time described by partial differential equations
- Amplitude, length, frequency, speed of the wave front

Marian Kupczynski

## Invisible versus Visible

From an airplane we notice a path in a jungle
- Was it done by loggers?
- Was it done by a twister?
- Was it done by a herd of elephants?

In a bubble chamber we notice a path
- Was it done by a moving point-like electron?
- Was it done by an extended proton?
- Was it done by a heavy ion?
- Was it done by some wave like phenomenon?

Einstein : a point-like particle has to be somewhere (QM is incomplete)  Bohr: we are not allowed to make space–time description of invisible details of quantum phenomena

Marian Kupczynski

## Properties

- An attributive property is a constant property of an object which can be observed and measured at any time and which is not modified by the measurement.
  Example: rest mass, electric charge

- A contextual property is a property revealed only in a specific experiments or under particular conditions and characterizes the interaction of the object with the external world or a measuring apparatus. Color of the chameleon, weight, spin projection and "probability"

Marian Kupczynski

## Common starting points: P=60\180=1\3

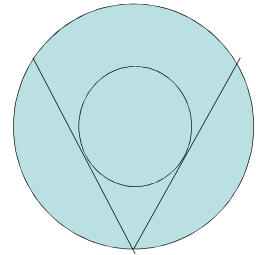

Marian Kupczynski

## Bertrand Paradox(1889)

Let us consider two concentric circles on a plane with radii R and R/2, respectively.

What is the probability P that a chord of the bigger circle chosen at random cuts the smaller one at least in one point?

VARIOUS ANSWERS POSSIBLE

Marian Kupczynski

## Chosen center: P=π( R/2)$^2$/π(R)$^2$=1/4

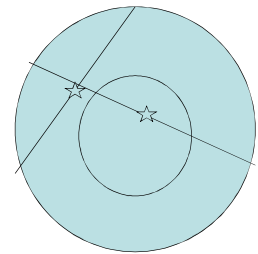

Marian Kupczynski

## Parallel sets: P=R\2R=1\2

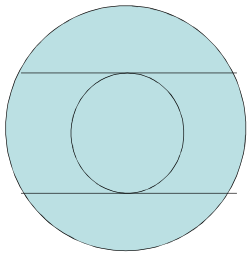

Marian Kupczynski

## Bertrand Paradox Solution

- A solution of the paradox is simple.
- The different values of P correspond to the different random experiments which may be performed in order to find the experimental answer to the Bertrand's question.
- Thus the probabilities have only a precise meaning, if the random experiments used for their estimation are specified.

Marian Kupczynski

## Frequentist estimation of probability

Flipping a fair coin N times:

HTTHHHTHTH……H

N=100  #(H)= 60  #(T)=40  p(H)~60/100=0.6

N=1000  #(H)= 450  #(T)=550    p(H)~45/100=0.45

N=100000  #(H)= 51000  #(T)=49000   p(H)~51/100=0.51

N tends to infinity→ P(H)=1/2

Marian Kupczynski

## Typical Experimental Set-UP II

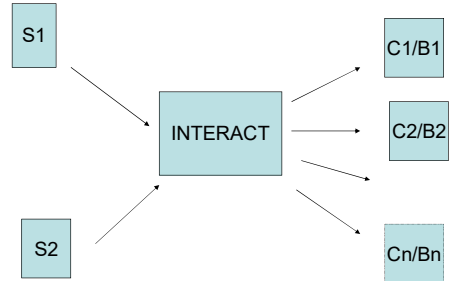

Marian Kupczynski

## Probability is a "contextual property" of a random experiment

The probability $p_{ij}$ is neither a property of the coin nor the property of the flipping device $D_j$. It characterizes only a particular random experiment: Flipping $C_i$ with a device $D_j$.

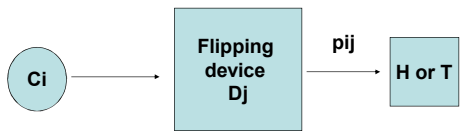

Marian Kupczynski

## Typical Experimental Set-UP III

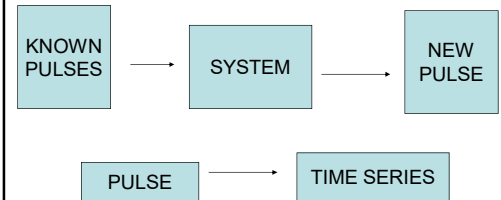

Marian Kupczynski

## Typical Experimental Set-UP I

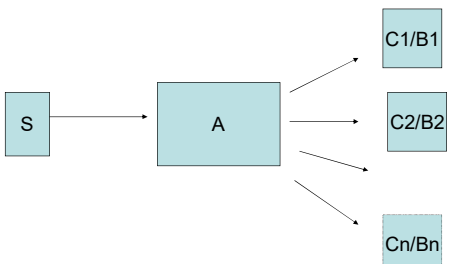

Marian Kupczynski

## Quantum Theory

- The theory provides abstract algorithms allowing to explain in quantitative way various physical phenomena.

- S - matrix description of scattering processes of elementary particles

Marian Kupczynski

## ALSO TRUE FOR

- Stochastic models describing the time evolution of trapped molecules, atoms or ions.
- There is no deterministic prediction for a single measurement or for a single time-series of events observed for a trapped ultra cold ions and quantum dots

Marian Kupczynski

## Experimental contexts

In QM one has conditional probabilities:
$P(A=a)=P(A=a|C)$ where C is a context of the experiment in which A is measured

Pure state preparation: ψ vector in Hilbert space

Observable measured **A**: Â self adjoint operator

Experimental setup to measure **A**: A

$E(A) = E(A|C) = E(A|\psi, A) = (\psi, \hat{A}\psi)$

Marian Kupczynski

## Statistical Interpretation of QM

IDENTICAL STATE PREPARATIONS REPEATED

Ensemble of Prepared Physical Systems → Wave Function or Density Matrix

Marian Kupczynski

## Statistical Interpretation SCI

QT GIVES STATISTICAL PREDICTIONS FOR:

- distribution of the results obtained in long runs of one experiment
- distribution of the results for several repetitions of the same experiment on a single system
- QT is an abstract statistical and contextual theory
- Statistical and contextual interpretation of QT will be called in this series of talks SCI

Einstein, Fock, Blokhintsev, Landau, Ballentine,, Bell, Theo, Andre, Luigi, Hans, Louis, Ana, Gerhard ,MK. ….

Marian Kupczynski

## Measurement of an observable O on an ensemble prepared in a pure state

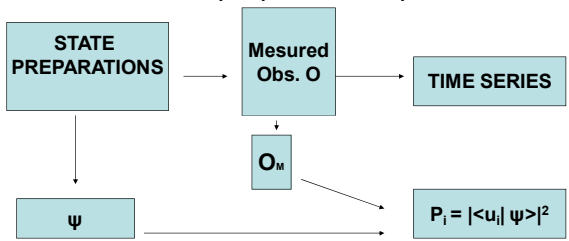

where $O_M$ is a Hermitian operator representing a measured physical observable O : $O_M u_i = a_i u_i$, $P_i = P(O=a_i)$

Marian Kupczynski

## Justified Questions

- Can we deduce quantum probabilities from some underlying more detailed theory of sub-phenomena?
- How in what sense may QT provide a complete description of individual physical systems?
- Does QT provide a complete description of experimental data? Is it predictably complete?

Marian Kupczynski

## Copenhagen Orthodoxy 1935

INDIVIDUAL PHYSICAL SYSTEM: t = T

Pure State → Unique Wave Function

**Instantaneous wave function reduction:**

**Any measurement causes a physical system to jump into one of the eigenstates of the dynamical variable that is being measured .**

Marian Kupczynski
EMQ15

## EPR PAPER (1935)

EPR : Two systems I +II in a pure quantum state which interacted in the past and which separated.

- A single measurement performed on one of the systems, for example on the system II , gives instantaneous knowledge of the wave function of the system I moving freely far away.

- By choosing two different incompatible observables to be measured on the system II it is possible to assign two different wave functions to the same physical reality ( the second system after the interaction with the first).

Marian Kupczynski

## QM postulates details

- S1: Any pure state of system is described by a unique wave function = a vector on a Hilbert space

- S2: Any observable is represented by a particular self adjoint operator

- S3: Any result of measurement of an observable is one of its eigenvalues

- S4: Any measurement causes the system to jump into one of the eigenstates of the dynamical variable that is being measured

- S5: If a jump occurs only the probabilities of obtaining particular experimental result can be calculated in the theory.

Marian Kupczynski

## QT:Contextual description

- Let us consider two experiments in which on the ensemble of EPR pairs prepared in a pure state $\Psi$ we measure different pairs of physical observables:
  In experiment 1: observables C and A
  In experiment 2: observables D and B.

- To describe the experiment 1 we develop a state vector $\Psi = \sum_i w_i |f_i\rangle |g_i\rangle$ using a basis formed by the eigenvectors of the operators $(\hat{C}, \hat{A})$ representing the observables C and A

- To describe the experiment 2 we develop a state vector $\Psi = \sum_j t_j |u_j\rangle |v_j\rangle$ using a basis formed by the eigenvectors of the operators $(\hat{D}, \hat{B})$ representing the observables D and B

Marian Kupczynski

## Postulates

- S6: We can't measure simultaneously complementary observables represented by non-commuting operators.

- S7: If we have two free physical systems then a wave function describing a pure state of a couple is a product of two independent wave functions each describing one member of a couple

- S8: State vector provides a complete description of the individual physical system

S1+S4+S7+S8 ⟶ EPR PARADOX

Marian Kupczynski

## EPR PRECISE FORMULATION

$$(1)\, \Psi = \sum_i w_i |f_i\rangle |g_i\rangle \to |f_r\rangle \quad if \quad A = a_r$$
$$where \quad \hat{C}|f_i\rangle = c_i|f_i\rangle \quad and \quad \hat{A}|g_i\rangle = a_i|g_i\rangle$$

$$(2)\, \Psi = \sum_j t_j |u_j\rangle |v_j\rangle \to |u_s\rangle \quad if \quad B = b_s$$
$$where \quad \hat{D}|u_j\rangle = d_j|u_j\rangle \quad and \quad \hat{B}|v_j\rangle = b_j|v_j\rangle$$

$$(1) \Leftrightarrow (\Psi, \hat{C}, \hat{A}) \qquad (2) \Leftrightarrow (\Psi, \hat{D}, \hat{B})$$

Marian Kupczynski

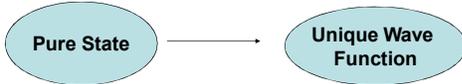

## EPR - explained

- EPR → two different wave functions assigned to the same physical reality ( the second system after the interaction with the first).
- Einstein (1936): Ψ function does not, in any sense, describe the state of one single physical system. Reduced wave functions describe different sub-ensembles of the systems.
- Ballentine (1998)…the habit of considering an individual particle to have its own wave function is hard to break.... though it has been demonstrated strictly incorrect ..

Marian Kupczynski

## BOHR

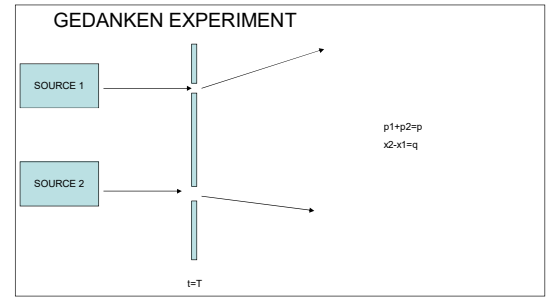

GEDANKEN EXPERIMENT

SOURCE 1

SOURCE 2

p1+p2=p
x2−x1=q

t=T

Marian Kupczynski

## EPR – example

Consider two systems I+II such that for t=T :
1) $p_1+p_2 = p$ and $x_2 - x_1 = q$
2) If we measure $p_2$ on the system II we get immediately
$$p_1 = p - p_2$$
3) If we measure $x_2$ on the system II we get immediately :
$$x_1 = x_2 - q$$

Without in any way disturbing the system I we can find a value of its linear momentum or of its position at the same moment of time with arbitrary precision.

**Quantum Mechanics is INCOMPLETE**

Marian Kupczynski

## BOHR's ANSWER

- A single couple of point like particles of known total momentum collides simultaneously at t=T with a two slit diaphragm, passes through the slits and a single measurement is performed at the end on one of these particles.
- To be able to get information about linear momentum $p_2$ of the second particle from the measurement of the linear momentum $p_1$ of the first particle we have to use a moving diaphragm.
- To get a precise information about the position $x_2$ of the second particle from the measurement of the position $x_1$ of the first particle the diaphragm has to be fixed.

Marian Kupczynski

## EPR-Local Realism-CDF

- Therefore it is justified to use a mental picture of a physical system for which it makes sense that it has at a given moment of time a definite position and a linear momentum
- Consequently the space time description of the sub phenomena in the micro world should not be abandoned
  **Quantum Mechanics is INCOMPLETE**

BOHR: One has to analyze the state preparation and experiments allowing to deduce $p_1$ or $x_1$ from the measurements of $p_2$ and $x_2$

Marian Kupczynski

## BOHR's Complementarity-Contextuality

- EPR inference requires different incompatible, but complementary, experiments.
- Bohr quickly noticed a problem with his arguments: if we start talking about point-like particles they have to be somewhere even if we do not observe them!
- He adopted then the idea of wholeness of quantum phenomena. *Strictly speaking , the mathematical formalism of quantum mechanics and electrodynamics merely offers rules of calculation for the deduction of expectations pertaining to observations obtained under well-defined experimental conditions specified by classical physical concepts*

Marian Kupczynski

## Statistical Interpretation (SCI)

IDENTICAL STATE PREPARATIONS REPEATED

Ensemble of Prepared Physical Systems → Wave Function or ρ

**A state vector or a density matrix describes an ensemble of the identical preparations of the physical systems.**

Marian Kupczynski
EMQ15

## CRITICISM ( inspired by Bohr 1935)

- In b) by saying that the particles had a time to separate we assume a mental image of two point-like particles which are produced and which after some time become separated and free. (Incorrect picture for a photon pairs?)

- We do not see any particular couple of the particles and we do not follow their space time evolution. We record only the clicks on the far away coincidence counters.

- To be able to deduce the value of a particular spin projection for the particle #2 from the measurement made on the particle #1 we need more information about each individual couple of the particles than we have in a simple coincidence experiment. MK

Marian Kupczynski

## EPR-BOHM (1951) SINGLET STATE)

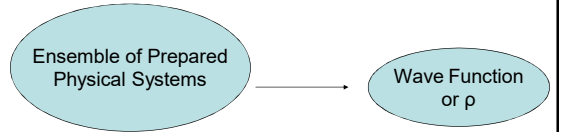

where $|+>_{\hat{p}}$ and $|->_{\hat{p}}$ are state vectors corresponding to the particle states in which the spin is "up" or "down" in the direction of $\hat{p}$ respectively.

a) Even if the orbital state is not stationary, the interactions do not involve spin and so the spin states will not change.

b) The particles are allowed to separate, and when they are well beyond the range of the interaction we can measure a component of spin of the particle #1 on the direction **p**.

Marian Kupczynski

## NO EPR-B PARADOX IN SCI

- A value of a contextual physical observable such as a spin projection resumes only the information about the interaction of the system with the measuring apparatus.
- A reduced one particle state $|+>_P$ describes the whole ensemble of the partners of the particles which were analyzed to have "spin down" by the analyzer P.
- For various directions **P** it is a different ensemble
- Quantum state reduction is not instantaneous!
- No Quantum magic!
- Bell 64:  Correlations : CRY FOR EXPLANATION!

Marian Kupczynski

## EPR-B PARADOX

c) Because the total spin is zero, we can predict with certainty, and without in any way disturbing the second particle, that the **p**- component of spin of particle #2 must have the opposite value.

d) Thus the **p**- component of spin of particle #2 is an element of reality for the particle #2 simultaneously for any direction **p**. CDF assumption in EPR-B

This contradicts QM!

Spin projection is discovered only when measured. Electron has to be somewhere! MK

Marian Kupczynski

## LOCALITY

Locality (Einsteinian separability)

We know that: distance=speed x time

Einstein: Maximal speed of physical influences is equal to speed of light in vacuum : c=300 000km/s

A ----------------------------------------------- B

What happens in A at $t_A$ can't influence what happens in B at $t_B$ if

$t_B - t_A ≤$ (Distance between A and B)/c

A space shuttle is now at the distance of $9 \times 10^7$ km. We receive a signal when was it sent?

Marian Kupczynski

## CAUSALITY

Bomb exploded on the plane and all passengers are dead.

The cause of their death was the explosion of the bomb

We put a seed in the ground and plant grows.

We take a poison and we die.

We switch off the light and it is dark in the room

We have regular high tides because of the gravitational attraction by the MOON .

In general: causal chain: E1→ E2→…→En

Marian Kupczynski

---

## Nonlocal randomness vs Physical Reality

- Bar-tailed Godwit (*Limosa lapponica baueri*) do the fall migration from Alaska to New Zealand in one hop.
- They fly 11 000 km in about eight days over the open Pacific Ocean, without stopping to rest or refuel.
- How could they do it if nonlocal randomness was a law of Nature?

Marian Kupczynski

---

## Reality versus solipsism

A table exists independently whether we measure its dimensions or not , whether we look at it or not.

We see a new track in the bubble chamber it was something real which made it happen.

We hear a click on the detector there is something real what made it happen even if we cannot see it and visualize it.

There are some real stuff behind the scenes in quantum phenomena.

Solipsism :A world which we perceive exists only in our imagination. Our theory describes only our beliefs and perceptions and not objective regularities existing in Nature.

Marian Kupczynski

---

## Local realism versus contextuality

- Local realism or counterfactual definiteness (CDF): measuring devices register values of physical observables existing independently whether they are measured or not.
- Contextuality: the values of contextual physical observables such as a spin projections are created in the interaction of the physical system with the measuring apparatus and they do not exist before the measurement

Marian Kupczynski

---

## Physical reality

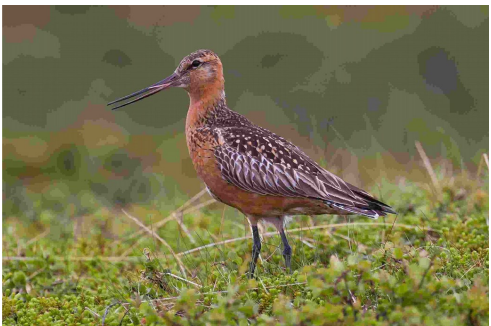

Marian Kupczynski

---

## Random variables and probabilistic models

1. Flipping a coin. We observe two possible events : H or T which are equally likely to occur. Following Kolmogorov we define a probability space of experiment Ω={H,T} and probability distribution p(H)=p(T)=1/2 called B(1/2)
2. We may associate the numbers to H and T in different ways using mappings called random variables. For example X1(H)=1 and X1(T)=0 or X2(H)=1 and X2(T)=2 and corresponding probability distributions P(X1=1)=P(X1=0)=1/2 and P(X2=1)=P(X2=2)=1/2
3. We define expectation value: $E(X) = \sum xp(x)$

   E(X1)= 0x1/2+1x1/2=1/2 , E(X2)=?
4. We define variance :Var(X)=E($X^2$)-E(X)$^2$=(σ(X))$^2$

Marian Kupczynski

## Covariance and stochastic independence

1. We flip two coins at the same time we observe HH,HT.. Now our probability space is $\Omega=(HH,HT,TH,TT)$ and we may introduce a multivariate random variable (X,Y) taking the values (1,1), (1,0), (0,1) and (0,0) and a joint probability distribution $p_{XY}(x,y) = p(X=x \text{ and } Y=y) = 1/4$
2. Since outcomes obtained on a coin 1 and 2 are independent we say that the random variables X and Y are stochastically independent:
   $p_{XY}(x,y) = p_X(x) p_Y(y)$ and $E(XY)=E(X)E(Y)$
   $$E(XY) = \sum xy p_{XY}(x,y) \quad E(X) = \sum x p_{XY}(x,y) \quad E(Y) = \sum y p_{XY}(x,y)$$
3. If $p_{XY}(x,y) \neq p_X(x) p_Y(y)$ then a covariance of (X,Y)
   $Cov(X,Y) = E(XY) - E(X)E(Y) \neq 0$

Marian Kupczynski

## PROBABILITY

- A probability is an objective contextual property of some random experiments

- In physics probability is not a subjective belief of some intelligent agents

## Finite samples and correlations

Let as assume that we measure the values of 2 random variables and Y and we obtain 2 sample of size n
X: $\{x_1, x_2, \ldots, x_n\}$ ; Y: $\{y_1, y_2, \ldots, y_n\}$;

Using these samples we may estimate expectations

$$E(X) \approx \bar{x} = (x_1 + \ldots + x_n)/n \quad E(Y) \approx \bar{y} = (y_1 + \ldots + y_n)/n$$

If we pair the outcomes: $(x_1,y_1), (x_2,y_2)\ldots$

$$E(XY) \approx corr(X,Y,n) = (x_1 y_1 + \ldots + x_n y_n)/n$$

Marian Kupczynski

## Intimate relation: Protocol → Probabilistic Model

1. Two draws with replacement : X= # of blue
   P (X=0)=1/9, P(X=1)=4/9, P(X=2)= 4/9
2. Two draws without replacement P(X=1)=2/3, P(X=2)=1/3

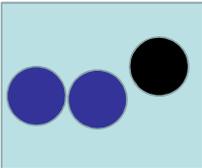

Marian Kupczynski

## Experiments and random variables

- Fair Coin Flipping : Heads →X=1 , Tails →X=0
  P(X=1)=P(X=0)=1/2   X obeys Bernoulli(1/2)

- Fair Dice Rolling : Face →X=i, i=1,2,3,4,5 and 6
  P(X=i)=1/6        X obeys Bernoulli(1/6)

The functions X are random variables!
Unfair coin or dice: Probability distribution unknown before gathering outcomes of many trials of the experiment.

## Joint Probability Distributions

1. Draw 1 with replacement, record the colour and the size :
RED →X1=1 , BIG →X2=1  P( X 1=i, X2=j ) where i,j=0 or 1.
Probability space Ω contains 4 elements: $|\Omega|=4$
2. Same experiment but record only the colour. Marginal probability:
P(X1=i)=P(X1=i, X2=1) +P(X1=i, X2=0)  $|\Omega|=2$

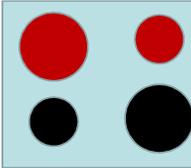

Marian Kupczynski

## Joint Probability Distribution

Each ball has two attributes: Colour and Size. ''Measurement'' of the colour does not change the size.

Colour and Size can be measured simultaneously or successively in any order.

Having the data for the experiment one does not need to measure the colour or the size separately because the marginal probabilities P(X1=i) and P(X2=j) can be estimated using the empirical joint probability distributions.

## Quantum Filters

$F_i$ →CREATOR of a contextual property "i" (passing by $F_i$ )

1) $F_i$ idempotent device : $F_i F_i = F_i$
2) Incompatible filters: $F_i F_j \neq F_j F_i$
   A "particle" having a property "i" is passed by the filter $F_j$ with a probability $p_{ij}$ acquiring a new property "j" or it is stopped. If the properties are compatible pij=1
3) No Joint Probability Distribution for Incompatible Properties do exist.

## Attributive Joint Probability Distributions

- If we measure simultaneously the values of $n$ random variables $X_1, X_2 \ldots X_n$ then we describe the experiment by a multivariate random variable $X=(X_1, X_2 \ldots X_n)$, a common probability space $\Lambda$, and a joint probability density function $p(x_1, \ldots, x_n)$.
- From $p(x_1, \ldots, x_n)$ by integrating over $(n-1)$ variables we obtain $n$ marginal probability density functions $p_{Xi}(x_i)$ describing n different random experiments
- For n=2

$$p_{X_1}(x_1) = \int_\Lambda p_{X_1 X_2}(x_1, x_2) d(x_2)$$

Marian Kupczynski

## Common cause and distant correlations.

- Let us consider two distant experiments **x** and **y** run by Alice and BOB in distant locations. They analyse the properties of binary signals 0 or 1 sent to them by Harry.
- Alice and BOB register time series of outcomes 010001…
- These time series may be correlated, if Harry prepares them to be correlated (Common cause) or uncorrelated if he chooses the bits sent to Alice and Bob by flipping two fair coins.
- Year by year increase of the salaries of teachers in France is correlated to the increase of price of apartments in Joao Pessoa due to the global inflation. (Common cause)

Marian Kupczynski

## Classical Filters- Macroselectors

Macroscopic objects → attributes $a_i$ i=1,2…n

$F_i$ filter selecting objects having the attribute $a_i$

1) $F_i$ idempotent device : $F_i F_i = F_i$
2) All filters are compatible: $F_i F_j = F_j F_i$
3) If a particle has a property $a_i$ is passed by a filter $F_i$ otherwise it is stopped.
4) In principle one can measure simultaneously all attributive properties on members of some classical statistical population described by some joint probability distribution.

## Why distant correlations?

- The correlations do not prove any causal relation between x and y .
- No communication or direct influence between x and y is needed for their existence.
- They may be due to a common cause or accidental.
- To describe them we need generalised joint probability distributions.

Marian Kupczynski

## Generalised joint probability distributions (GJPD)

2 random experiments x and y to measure A and B

two samples : $\{a_1, a_2, \ldots a_n \ldots\}$ and $\{b_1, b_2, \ldots b_n \ldots\}$

math. stat.: observations of two time series of random variables $\{A_1, A_2, \ldots A_n \ldots\}$ and $\{B_1, B_2, B_n \ldots\}$. To study correlations we need to find joint probability distribution

Problem how to pair the data in order to find a GJPD?
IN GENERAL: NON UNIQUE SOLUTION!

GJPD depends on a protocol how the pairing is done. *In SPCE: width of the time windows, coincidence technique etc.*

Marian Kupczynski

## EPR-B experiment with electrons

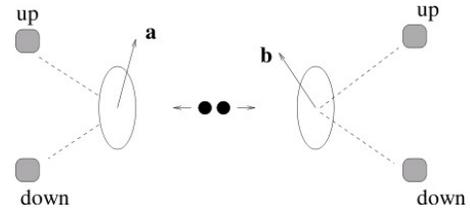

Marian Kupczynski

## Correlations depend on Pairing

- Pairing $S_{1k}$ : { $(a_1, b_k), (a_2, b_{k+1}), (a_3, b_{k+2})\ldots$}
- Pairing $S_R$ : { pairs $(a_s, b_t)$ chosen at random}

Example $S_1 = 01010101..$  $S_2 = 101010101..$
Pairing $S_{1k}$  k=odd   strong anti-correlation
Pairing $S_{1k}$  k= even   strong correlation
Pairing $S_R$   no correlation!

Marian Kupczynski

## Strong correlations of measurement outcomes in distant laboratories and CDF

Let us consider an idealized EPR-B experiment with electrons:

- A source is sending n pairs of electrons, one electron is sent to Alice and its "partner" is sent to Bob in distant location
- A measurement of spin projections are performed at various directions characterized by angles $\theta_A$ and $\theta_B$ respectively. Registered results are either spin-up (+) or spin-down (−) with respect to chosen axis.
- Due to the conservation of angular momentum in the singlet state, the sum of the spin projections should vanish for <u>any</u> chosen angle
- Prediction of QM :

$$E(a,b) = -a \cdot b = -\cos(\theta_A - \theta_B)$$

Marian Kupczynski

## Correlations

- Once the pairing is defined GJPD and correlations may be estimated.

- In general GJPD do not factorize:

$P(a, b \mid x, y) = P(A=a, B=b \mid x, y, S_1, S_2) \neq P(A=a \mid x, S_1) P(B=b \mid y, S_2)$

Marian Kupczynski

## Estimation of expectation values (Wikipedia)

| Anti-parallel | Pair 1 | Pair 2 | Pair 3 | Pair 4 | ... | Pair n | |
|---|---|---|---|---|---|---|---|
| Alice, 0° | + | − | + | + | ... | − | |
| Bob, 180° | + | − | + | + | ... | − | |
| Correlation = | ( +1 | +1 | +1 | +1 | ... | +1 | ) / n = +1 |

| Parallel | Pair 1 | Pair 2 | Pair 3 | Pair 4 | ... | Pair n | |
|---|---|---|---|---|---|---|---|
| Alice, 0° | + | − | + | − | ... | + | |
| Bob, 0° or 360° | − | + | + | − | ... | − | |
| Correlation = | ( -1 | -1 | -1 | -1 | ... | -1 | ) / n = -1 |

$$C(a,b,n) = (N(+,+) + N(-,-) - N(+,-) - N(-,+))/n$$

Marian Kupczynski

## Puzzle of distant EPR-B correlations

- Bell 1964 : Strong correlations In EPR-B experiments cry for an intuitive explanation ! I AGREE!
- If we assume that the values of spin projections in all directions are predetermined by a source (CDF) and registered passively by detectors then one may prove inequalities violated by predictions of QM !
  IMPORTANT RESULT!

Marian Kupczynski

## Bell Inequalities : proof 1964

- If we assume strict anti-correlations $A(a,\lambda)=-B(a,\lambda)=\pm 1$ thus:

$$E(a,b) = -\int_\Lambda A(a,\lambda)A(b,\lambda)\rho(\lambda)d\lambda$$

- We obtain first Bell inequalities

$$|E(a,b) - E(a,c)| = |-\int_\Lambda [A(a,\lambda)A(b,\lambda) - A(a,\lambda)A(c,\lambda)]\rho(\lambda)d\lambda| =$$

$$= |\int_\Lambda A(a,\lambda)A(b,\lambda)[A(b,\lambda)A(c,\lambda)-1]\rho(\lambda)d\lambda| \leq 1 + E(b,c)$$

- These inequalities for some choices of a, b and c are violated by the correlations predicted by QM

Marian Kupczynski

## Local realistic hidden variable model: LRHV ( Bell 64)

- A source is producing a statistical ensemble of electron pairs labelled by a ("hidden") parameter $\lambda \in \Lambda$ and subject to some probability distribution $\rho(\lambda)$. Values registered by Alice or Bob are functions of $\lambda$ and of local orientation vectors **a** and **b** of Stern-Gerlach magnets : $A(\mathbf{a},\lambda)=\pm 1$ and $B(\mathbf{b},\lambda)=\pm 1$.
- Correlations are described by expectation values :

(1) $\quad E(a,b) = \int_\Lambda A(a,\lambda)B(b,\lambda)\rho(\lambda)d\lambda$

where a denotes **a** and b denotes **b**. $C(a,b,n) \approx E(a,b)$ for large $n$.

Marian Kupczynski

## Proof of CHSH inequality

1. First:

$$|E(a,b) - E(a,b')| = |\int_\Lambda A(a,\lambda)B(b,\lambda)[1 \pm A(a',\lambda)B(b',\lambda)]\rho(\lambda)d\lambda -$$

$$-\int_\Lambda A(a,\lambda)B(b',\lambda)[1 \pm A(a',\lambda)B(b,\lambda)]\rho(\lambda)d\lambda | \leq$$

$$\leq \int_\Lambda [1 \pm A(a',\lambda)B(b',\lambda)]\rho(\lambda)d\lambda + \int_\Lambda [1 \pm A(a',\lambda)B(b,\lambda)]\rho(\lambda)d\lambda = 2 \pm (E(a',b') + E(a',b))$$

2. From 1 we get $|E(a,b) - E(a,b')| + |E(a',b') + E(a',b)| \leq 2$

3. Then $S(a,b,a',b') = E(a',b) + E(a',b') + E(a,b) - E(a,b') \leq 2$

4. Interchanging $a$ with $a'$ we obtain a standard form of CHSH
   $S = E(a,b) + E(a,b') + E(a',b) - E(a',b') \leq 2$

5. Simpler and more general proofs of S≤2 will be given in next talks

Marian Kupczynski

## CONFLICT with QM

(1) $\quad E(a,b) = \int_\Lambda A(a,\lambda)B(b,\lambda)\rho(\lambda)d\lambda$

QM prediction :

(2) $\quad E(a,b) = -a\bullet b = -\cos(\theta_A - \theta_B)$

Using (1) one may not reproduce (2) for all angles.

Marian Kupczynski

## Clauser-Horne Model :1974

- A source emits pairs of correlated particles labelled by a ("hidden") parameter $\lambda \in \Lambda$ and subject to some probability distribution $\rho(\lambda)$.
- Each particle has to pass through an analyser with an orientation vector ( $a$ or $b$ ) before reaching one of the detectors (D1 or D2).
- Each particle from a pair $\lambda$ has a probability $p_1(a, \lambda)$ or $p_2(b, \lambda)$ to be registered by the detector after passage by analyser ($a$ or $b$).
- Evoking locality Clauser - Horne postulate:

(!) $\quad p_{12}(a, b,\lambda) = p_1(a, \lambda)\ p_2(b, \lambda)$

they define various detection probabilities $P_1(a)$, $P_2(b)$ and $P_{12}(a,b)$ and find CH inequality which may be compared with the predictions of QM.

$$P_1(a) = \int_\Lambda p_1(a,\lambda)\rho(\lambda)d\lambda, \quad P_2(b) = \int_\Lambda p_2(b,\lambda)\rho(\lambda)d\lambda$$

$$P_{12}(a,b) = \int_\Lambda p_1(a,\lambda)p_2(b,\lambda)\rho(\lambda)d\lambda$$

It is more exact to call (!) a stochastic independence for a fixed lambda

Marian Kupczynski

## CH inequality

- Using CH inequality in the form below it is difficult to find contradiction with the prediction of QM in realistic experiments

$$\frac{P_{12}(a,b) - P_{12}(a,b') + P_{12}(a',b) + P_{12}(a',b')}{P_1(a') + P_2(b)} \leq 1$$

- However various variants of finite sample CH and Eberhard inequalities were used in recent tests of local realism.

The model proposed first by CH led to the formulation of modern stochastic hidden variable models (SHVM) which will be defined later in this talk.

Several ingenious spin polarization correlation experiments (SPCE) reported the violation of Bell, CHSH, CH.,,inequalities.

Marian Kupczynski

## QT→No strict anti-correlations!

- If $A$ and $B$ are the spin projections on two directions characterized by angles $\theta_A$ and $\theta_B$ respectively then for a perfect singlet state :

$$E(AB|\psi) = -\cos(\theta_A - \theta_B)$$

- No sharp directions exist in Nature. Thus QT prediction is :

$$E(AB|\psi) = -\iint_{I_A I_B} \cos(\theta_1 - \theta_2) d\rho_A(\theta_1) d\rho_B(\theta_2)$$

where $I_A$ and $I_B$ are small intervals around $\theta_A$ and $\theta_B$
(MK, Phys.Lett. 121,51-53,1987)

Marian Kupczynski
EMQ15

## Spin polarization correlation experiments SPCE

- A pulse of laser hitting the non linear crystal produces two correlated signals propagating in opposite directions.
- Each of these fields have a property that they produce a clicks when hitting detectors.
- When we place analyzers in front of the detectors after interaction of the fields with the analyzers we obtain two time series of clicks on the far away detectors which are strongly correlated.

Marian Kupczynski

## Mixed quantum states produced in SPCE

- Werner State: $\rho = (V|\Psi\rangle\langle\Psi| + (1-V)I)/4$
- Eberhard state:
- Köfler et al.: $\rho_r = (1+r^2)^{-1/2}(|H\rangle\langle V| + r|V\rangle\langle H|)$

$$\rho_r = \frac{1}{1+r^2}\begin{bmatrix} 0 & 0 & 0 & 0 \\ 0 & 1 & Vr & 0 \\ 0 & Vr & r^2 & 0 \\ 0 & 0 & 0 & 0 \end{bmatrix}$$

1>r>0   1>V but in fact V is complex and no zero elements in ρ .(r=0.297,V=0.965 good fit of the data)

Marian Kupczynski
EMQ15

## EPR-B-with photons in maximally entangled state

$$\Psi = (|H\rangle \otimes |V\rangle + |V\rangle \otimes |H\rangle)/\sqrt{2}$$

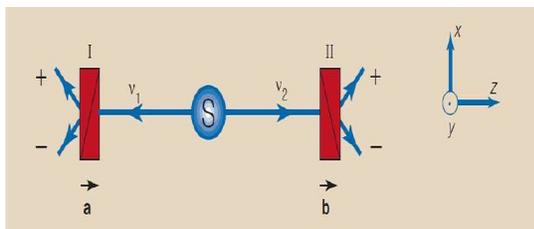

Marian Kupczynski

## Why distant correlations?

- The correlations do not prove any causal relation between x and y .

- No communication or direct influence between x and y is needed for their existence.

Marian Kupczynski

## Generalised joint probability distributions (GJPD)

2 random experiments x and y to measure A and B

two samples : $\{a_1, a_2, \ldots a_n \ldots\}$ and $\{b_1, b_2, \ldots b_n \ldots\}$

math. stat.: observations of two time series of random variables $\{A_1, A_2, \ldots A_n \ldots\}$ and $\{B_1, B_2, B_n \ldots\}$. To study correlations we need to find joint probability distribution

Problem how to pair the data in order to find a GJPD?
   IN GENERAL: NON UNIQUE SOLUTION!

GJPD depends on a protocol how the pairing is done. *In SPCE :width of the time windows, coincidence technique etc.*

Marian Kupczynski

---

## SPCE – STRONG CORRELATIONS

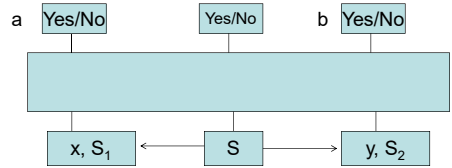

2 correlated signals $S_1$ and $S_2$ produced by a source S are hitting the measuring devices x and y producing correlated outcomes $a = \pm 1$ or 0 and $b = \pm 1$ or 0. Zero means no click detected.

Marian Kupczynski

---

## Correlations depend on Pairing

- Pairing $S_{1k}$ : $\{(a_1, b_k), (a_2, b_{k+1}), (a_3, b_{k+2}) \ldots\}$
- Pairing $S_R$ : $\{$ pairs $(a_s, b_t)$ chosen at random$\}$

Example $S_1 = 01010101..$ $S_2 = 101010101..$
Pairing $S_{1k}$ k=odd       strong anti-correlation
Pairing $S_{1k}$ k= even     strong correlation
Pairing $S_R$    no correlation!

Marian Kupczynski

---

## Bell 64 and CH HIDDEN VARIABLES INTRODUCED

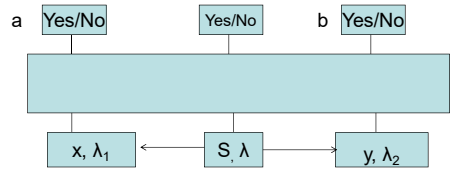

a is underlined{determined} locally by a setting label x and the values of $\lambda_1$ describing the signal $S_1$ in a moment of measurement. In a similar way b is determined by y and $\lambda_2$

Marian Kupczynski

---

## Correlations

- Once the pairing is defined GJPD and correlations may be estimated.

- In general GJPD do not factorize:
$P(a, b | x, y) = P(A=a, B=b|x, y, S_1, S_2) \neq P(A=a|x, S_1) P(B=b|y, S_2)$

Marian Kupczynski

---

## Probabilistic models

Outcome a is a value of a random observable A and b a value of B. All experiments are described using the same probability space $\Lambda$ and the same probability distribution $P((\lambda_1, \lambda_2))$ for all the settings (x,y)

- Bell 64: LRHVM or *Bertlmann's Socks Model*.

$$E(AB) = \sum_{\lambda \in \Lambda} P(\lambda_1, \lambda_2) A(\lambda_1) B(\lambda_2)$$

- CH 74: SHVM or *Pair of Dice Model*

$$E(AB | x, y) = \sum_{\lambda \in \Lambda} P(\lambda_1, \lambda_2) E(A | x, \lambda_1) E(B | y, \lambda_2)$$

$$P(a, b | x, y) = \sum_{\lambda \in \Lambda} P(\lambda_1, \lambda_2) P(a | x, \lambda_1) P(b | y, \lambda_2)$$

Marian Kupczynski

## Classical local hidden variable models

- In local realistic hidden variable models (LHRVM) **BELL 64** signals are represented as an ensemble of correlated photon pairs. labelled by $(\lambda_1, \lambda_2)$. having predetermined spin projections on different axes of polarization analyzers.

- In stochastic hidden variable models (SHVM) measurement outcomes on members of a given pair ,labelled by $(\lambda_1, \lambda_2)$ ,are produced at random and are stochastically independent! (also incorrect protocol)
- Hybrid Bell 1971 model (incorrect protocol)·

Marian Kupczynski

## STRONG CORRELATIONS EXPLAINED

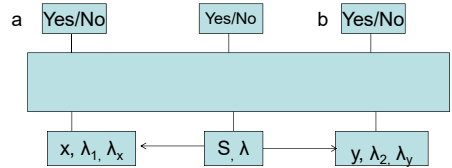

a=±1 or 0 is determined locally by the values of $\lambda_1$ and $\lambda_x$ describing the signal $S_1$ and the measuring device x respectively. Similarly b =±1 or 0 is determined by y and $\lambda_y$.

Marian Kupczynski

## Bell model 1971, Not perfectly correlated outcomes

- A source is sending a statistical ensemble of particle pairs described by ("hidden") parameters $\lambda \in \Lambda$ and subject to some probability distribution $\rho(\lambda)$.
- The instruments can contain hidden variables.
- For each pair Alice and Bob register one of three events coded as follows

    A(**a**, λ,..) =1 ; B(**b**, λ,..)=1   ('spin up' detector triggered)
    A(**a**, λ,..)=-1;  B(**b**, λ,..) =-1 ('spin down' detector triggered)
    A(**a**, λ,..) = 0 ; B(**b**, λ,..) =0   (particle undetected)

- Now Bell introduces the averages $|\bar{A}_a(\lambda)| \leq 1$ and $|\bar{B}_b(\lambda)| \leq 1$

(!) $$E(a,b) = \int_\Lambda \bar{A}_a(\lambda)\bar{B}_b(\lambda)\rho(\lambda)d\lambda$$

- Using (!) one proves immediately CHSH. Bell says that the averages are taken over instrument variables. The model (!) does not describe realizable coincidence experiments (SPCE).

Marian Kupczynski

## Meaning of hidden variables

Properties of correlated signals arriving to PBS-D module at time t, as perceived by them , are coded by $(\lambda_1, \lambda_2)$. Microstates of PBS-D modules at time t are coded by $(\lambda_x, \lambda_y)$. In function of these parameters we observe in a locally causal way a click or a no-click on corresponding detectors what is coded by :

A$(\lambda_1, \lambda_x)$=0, ±1 and B $(\lambda_2, \lambda_y)$=0, ±1

If we change (x ,y) into $(x_1, y_1)$ the properties of the signals as perceived by them are different thus

$$P(\lambda) = P_{xy}(\lambda_1, \lambda_2)P_x(\lambda_x)P_y(\lambda_y)$$

## Contextuality-KOLMOGOROV

EACH RANDOM EXPERIMENT- HAS IT'S OWN

PROBABILITY SPACE $\Lambda$ CONTAINING ALL POSSIBLE OUTCOMES OF THE EXPERIMENT

## Local Probabilistic Model for SPCE

$$E(AB) = E(AB|x,y) = \sum_{\lambda \in \Lambda_{xy}} P(\lambda)A(\lambda_1, \lambda_x)B(\lambda_2, \lambda_y)$$

where A$(\lambda_1, \lambda_x)$=0, ±1 and B $(\lambda_2, \lambda_y)$=0, ±1

$$P(\lambda) = P_{xy}(\lambda_1, \lambda_2)P_x(\lambda_x)P_y(\lambda_y)$$

Now there is no common hidden parameter space $\Lambda$ and $\Lambda_{xy}$ are different hidden parameter spaces for each pair (x,y) .

It is impossible to prove B-CHSH and CH inequalities!

Marian Kupczynski

## CHSH-BELL PROOFS

THE EXISTENCE OF A COMMON PROBABILITY TAKEN FOR GRANTED

$$\Lambda_{xy} \neq \Lambda_{x'y} \neq \Lambda_{xy'} \neq \Lambda_{x'y'} \neq \Lambda$$

FATAL CONTEXTUALITY LOOPHOLE

Marian Kupczynski

## Simple inequality to prove CHSH

We start with four integer numbers A,,B,A',B' such that their absolute value is 1: |A|=|B|=|A'|=|B'|=1 then

$$|s| = |AB + AB' + A'B - A'B'| \leq$$
$$|A||B+B'| + |A'||B-B'| \leq$$
$$|B+B'| + |B-B'| \leq 2$$

If B=-B' the first term vanishes and **we get |2B|≤2**. If B=B' the second term vanishes and **we get |2B'|≤2**
The equation |s|≤2 is also valid if |A|≤1… |B'|≤1 …..

Marian Kupczynski

## It was noticed by :

Accardi, Fine, Hess, Khrennikov, M.K, Michielsen, de Muynck, Niewenhuizen, Pitovsky, Philipp, De Raedt,…

Vorob'ev (1962): 'Is it possible to construct always the joint probability distribution for any triple of only pairwise measurable observables?' NON.

Yes only if their expectation values satisfy Bell inequalities . Similar result was proven by BOOLE

Marian Kupczynski

## Simple proof of CHSH: Bell 64

We replace A by A($\lambda_1$); B by B($\lambda_2$); A' by A'($\lambda_1$) and B' by B'($\lambda_2$)

(1) $$E(AB) = \sum_{\lambda \in \Lambda} P(\lambda_1, \lambda_2) A(\lambda_1) B(\lambda_2)$$

$$|S| = |E(AB) + E(AB') + E(A'B) - E(A'B')| \leq$$
$$|\sum_{\lambda} P(\lambda_1, \lambda_2)| |s(\lambda_1, \lambda_2)| \leq 2$$

To obtain CHSH : |S|≤2 we must use the same probability space Λ to describe all different experimental settings (x, y)

Marian Kupczynski

## BI special case of Boole Inequalities

In 1862, Boole showed that whatever process generates a data set S of triples of variables (S1,S2,S3) where Si = ±1, then the averages of products of pairs SiSj in a data set S have to satisfy the equalities very similar to BI

Extensive review and generalization:H. De Raedt, K.Hess and K.Michielsen,"Extended Boole-Bell Inequalities Applicable to Quantum Theory", *J. Comp. Theor. Nanosci.* **8**, 1011-1039,(2011)

## CHSH valid in SHVM

If we replace A by E(A|x,λ1) and B by E(B|x,λ2) etc and use

$$E(AB|x,y) = \sum_{\lambda \in \Lambda} P(\lambda_1, \lambda_2) E(A|x, \lambda_1) E(B|y, \lambda_2)$$

to prove immediately CHSH inequality in SHVM

We may prove also CHSH in QM for a convex sum of separable states

Marian Kupczynski

## QT Contextual description

Density operator ρ for a pure state: $\rho = |\Psi\rangle\langle\Psi|$

Local Observables: $\hat{A}_1 = \hat{A} \otimes I \qquad \hat{B}_1 = I \otimes \hat{B}$

Conditional expectation values:

$$E(A|\psi) = Tr\rho\hat{A}_1 \qquad E(AB|\psi) = Tr\rho\hat{A}_1\hat{B}_1$$

Covariance:

$$\text{cov}(A,B|\rho) = E(AB|\rho) - E(A|\rho)E(B|\rho)$$

Density operator ρ for a convex sum of separable states:

$$\rho = \sum_{i=1}^{k} p_i \rho_i \otimes \tilde{\rho}_i$$

## SHVM → Impossible Experimental Protocol

- Repeat the experiment x on a first photon from a pair ($\lambda_1, \lambda_2$) several times and find an estimate of the probability P(a | x, $\lambda_1$). Repeat the experiment y on a second photon from a pair find an estimate of P(b |y, $\lambda_2$) (IMPOSSIBLE TO IMPLEMENT IN SPCE)
- Find an estimate of:
  P(a , b | x ,y ,λ)= P(a | x,$\lambda_1$) P(b |y,$\lambda_2$)
- Repeat the steps 1 and 2 for the next pair of photons
- Average the data obtained after many repetitions of the preceding steps and obtain an estimate of
  P(a , b | x ,y )

## Pairs of free distant physical systems who never interacted → CHSH

Density matrix → convex sum of separable states:

$$E(AB|\rho) = \sum_{i=1}^{k} p_i E(A|\rho_i) E(B|\tilde{\rho}_i)$$

if local expectation values are:

$|E(A|\rho_i)| \leq 1 \qquad |E(B|\tilde{\rho}_i)| \leq 1$

Using QT one easily proves: CHSH :

In this case QT predictions can be reproduced by a local stochastic hidden variable model (SHVM).

$$|E(AB|\rho) - E(AB'|\rho)| + |E(A'B|\rho) + E(A'B'|\rho)| \leq 2$$

## SHVM-Stochastic Independence

$$P(a,b|x,y) = \sum_{\lambda \in \Lambda} P(\lambda_1, \lambda_2) P(a|x,\lambda_1) P(b|y,\lambda_2)$$

(!!)     P(a , b |x ,y ,λ)= P(a |x ,$\lambda_1$) P(b |y ,$\lambda_2$)

(!!) is only true if :there is no correlation between random variables A and B for each fixed λ

Experimental Protocol

For each λ estimate P(A |a ,λ) and P(B |b ,λ) in all the directions and then average over all λ.

This protocol is impossible to implement in SPCE therefore the violation of CHSH and CH has NOTHING TO DO WITH NONLOCALITY OF NATURE.

Marian Kupczynski

## SPCE→QT→ Singlet State → NO B-CHSH

Each couple of experiments (x , y) is described by its own probability space and a probability distribution is determined by a triplet:

( $\rho = |\Psi\rangle\langle\Psi|$ , $\hat{A}_1 = \hat{A} \otimes I$ , $\hat{B}_1 = I \otimes \hat{B}$ )

There is no joint probability distribution of measured spin projections in all directions .

One cannot prove B-CHSH, CH etc and QT predictions violate these inequalities.

## JPD of measurable random variables

Probability space contains all possible outcomes of all measurable random variables.

By integrating joint probability distributions (JPD) over some variables one obtails a marginal probability distribution describing a realizable random experiment (in which not all compatible random variables are measured).

## JPD of hidden random variables

- Probabilistic model using joint probability distribution of hidden variables is only a mathematical construction which specifies a plausible invisible experimental protocol explaining how individual outcomes are obtained in a given random experiment.
- Integrating over some variables in a hidden variable model one does not obtain marginal probability distributions of some realizable random experiment. One only obtains a hidden variable model for a new experiment with an invisible experimental protocol which is impossible to implement.

Kupczynski, M., Bell Inequalities, Experimental Protocols and Contextuality, Found. Phys. (12 Dec 2014), doi:10.1007/s10701-014-9863-     (arXiv:1411.7085)

## Mysterious entanglement Schrödinger(1936)

No matter how far apart the particles are ………they are not really "free"→

Entanglement, Quantum Steering →

Mystery , spooky action at the distance etc???

NO,NO, NO!!!!

Marian Kupczynski

## Bell 71 , 85 : impossible experimental protocol

(1) $$E(AB) = E(AB|x,y) = \sum_{\lambda \in \Lambda_{xy}} P(\lambda) A(\lambda_1, \lambda_x) B(\lambda_2, \lambda_y)$$

is consistent with experimental protocol of SPCE. After the summation over $(\lambda_x, \lambda_y)$ we obtain

(2) $$E(AB|x,y) = \sum_{\Lambda_{12}} P(\lambda_1, \lambda_2) \overline{A}(\lambda_1) \overline{B}(\lambda_2)$$

where $\overline{A}(\lambda_1) = \sum_{\Lambda_x} A(\lambda_1, \lambda_x)$

Eq. 2 defines a protocol: for each couple estimate the averaged values of A and B next multiply them and average the outcomes for all pairs to estimate E(A,B|x,y). This cannot be done in SPCE!

## Correlations Explained

- Distant EPR-B correlations are imperfect !
- The signals at the moment of the measurement conserve a memory of the source.
- This memory is partially preserved in causal locally deterministic interactions with the measuring devices .
- Each setting is described by its own probability space thus it is enough parameters to fit any data.

Marian Kupczynski

## Meaning of BI,CHSH and CH violation

"An entangled pair of photons" resembles neither

"a pair of Bertlmann's socks"  nor

"a pair of fair dice".

Violation of Bell-type inequality proves that the results of measurements are neither predetermined nor obtained in irreducibly random way.It has nothing to do with the non-locality of Nature!.

## Contextual Causal Diagram

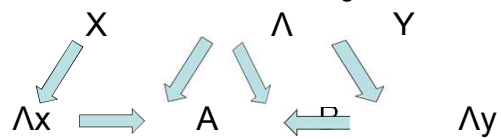

Experimenter has a free will: the settings (x, y) are chosen in any way he wants  but parameters Λx and Λy depend on settings (x , y) chosen.

Marian Kupczynski

## Lessons for QI community

"Once acquired, the habit of considering an individual particle to have its own wave function is hard to break.... though it has been demonstrated strictly incorrect"

L. Ballentine

According to statistical and contextual interpretation of QT:

1. A state vector is not an attribute of a qubit which may be changed instantaneously.
2. Correlations between outcomes obtained in distant locations can be explained in a local and causal way. No quantum magic is needed!
3. Unperformed experiments have no outcomes! (Peres)

Marian Kupczynski

## Can we discover that QT is emergent without constructing a specific sub-quantum model?
## YES!

Marian Kupczynski

## Jaynes

"He who confuses reality with his knowledge of reality generates needless artificial mysteries." (1989)

## Statistical interpretation of QM

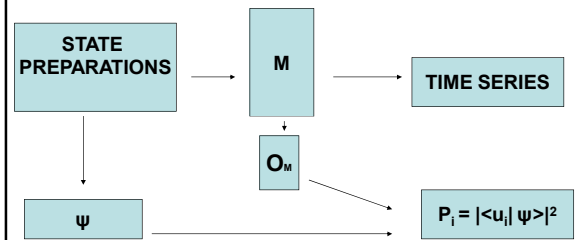

**Does it provide a complete description of the data?**

Marian Kupczynski

## SHVM INCORRECT THUS

VIOLATION OF B- CHSH

↓

NO IRREDUCIBLE RANDOMNESS

↓

EMERGENT QT → DESCRIPTION OF SUBPHENOMENA?

Marian Kupczynski

## Typical Experimental DATA

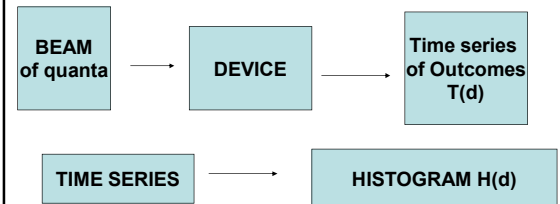

Marian Kupczynski

## Biased data analysis

- Time-series of experimental data were analyzed in terms of empirical probability distributions which could average out any stochastic fine structure of these time-series not predicted by QT.
- Simple statistical tests may be used to search for fine structures in time-series of data and check the predictive completeness and limitations of QT.

Marian Kupczynski

## Purity tests and completeness of QT

- Any model of sub-phenomena describes a pure quantum state as a mixed statistical ensemble with respect to some additional variables?
- If the mixture is not perfect then by changing the intensity or geometry of the beam we may obtain a sub-ensembles having slightly different properties than the initial ensemble?
- Using purity tests we may detect this effect therefore the purity tests may be used as the tests of the completeness of QM?
MK(1986,2002)

Marian Kupczynski

## Fine structure in time series

- Let us consider a random experiment which can give only two outcomes: 1 or -1. We repeat this experiment 2n times and we obtain a time series of the results:

    1,-1,1,-1,...,1,-1.

- By increasing the value of n the relative frequency of getting 1 can approach 1/2 as close as we wish. However it is not a complete description of the time series.

Marian Kupczynski

## PURITY TESTS

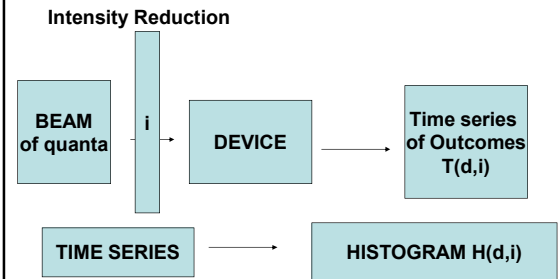

Marian Kupczynski

## Another fine structure

- Another example could be:
  1,-1,-1,-1,1,1,1,-1,-1,1,1,1,-1,1,-1,-1,-1

- The probability distribution does not provide the complete description of these time series of the data.

Marian Kupczynski

## Non parametric compatibility tests

THE STATISTICAL HYPOTHESIS $H_0$ TO TEST:

All the samples T(d,i) are drawn from the same unknown statistical population of the random variable X(d) associated to the observable measured by a device d.

An extensive use of the non-parametric statistical tests is needed.

Marian Kupczynski

## Is QT really predictably complete?

- If the answer is yes it means that the time series of the experimental data is completely described by the probability distribution given by QT?
- IT HAS TO BE TESTED AND NOT TAKEN FOR GRANTED?
- PURITY TESTS
- DETAILED STUDY OF TIME SERIES TO DETECT FINE STOCHASTIC STRUCTURE UPREDICTED BY QT!.

Marian Kupczynski

## Some papers on BI recommended by A. Khrennikov

Andrei Khrennikov (https://lnu.se/en/staff/andrei.khrennikov/)
- CHSH inequality: Quantum probabilities as classical conditional probabilities, arXiv:1406.4886 [quant-ph]
- Quantum probabilities and violation of CHSH-inequality from classical random signals and threshold type properly calibrated detectors, arXiv:1111.1907 [quant-ph]
- After Bell, arXiv:1603.08674 [quant-ph]
- Bell as the Copernicus of Probability, arXiv:1412.6987 [quant-ph]
- Quantum nonlocality or nonergodicity? A critical study of Bell's arguments ,arXiv:quant-ph/0512178
- Einstein and Bell, von Mises and Kolmogorov: reality and locality, frequency and probability, arXiv:quant-ph/0006016
- Khrennikov,, A. Probability and Randomness : Quantum versus Classical. ; Imperial College Press: London, England.2016

Marian Kupczynski

## Statistical inference and sample homogeneity loophole. Important closing remarks about statistical significance tests.

- Statistical significance tests are based on the assumption that trials are independent and that a sample is homogenuos, As we demonstrated recently with Hans de Raedt (Breakdown of statistical inference from some random experiments *Computer Physics Communications* **200**, 168 (2016)) sample inhomogeneity invalidates significance tests.
- If contextuality of quantum experiments is correctly taken into account then various Bell type inequalities cannot be proven therefore their violation in SPCE is not surprising.
- However we are not sure how significant was this violation because *sample homogeneity loophole* could not be closed . Namely the homogeneity of experimental samples was not or could not be tested.

Marian Kupczynski

## Some papers on BI recommended by H. De Raedt

Hans de Raedt :http://www.compphys.org/compphys0/publications.html

1. K. De Raedt, H. De Raedt, and K. Michielsen, "A computer program to simulate the Einstein-Podolsky-Rosen-Bohm experiment with photons", Comp. Phys. Comm. 176, 642 – 651 (2007).
http://rugth30.phys.rug.nl/pdf/EPJB-53-139-2006.pdf

2. S. Zhao, H. De Raedt, and K. Michielsen, "Event-by-event simulation model of Einstein-Podolsky-Rosen-Bohm experiments", Found. Phys. 38, 322 -347 (2008).
http://rugth30.phys.rug.nl/pdf/shu5.pdf

Proceedings of the previous Joao Pessoa summer school:
3. K. Michielsen and H. De Raedt, "Event-based simulation of quantum physics experiments", Int. J. Mod. Phys. C 25, 1430003 (2014).
http://rugth30.phys.rug.nl/pdf/IJMPC25.1430003.2014.pdf

4. H. De Raedt, K. Hess, and K. Michielsen,
"Extended Boole-Bell inequalities applicable to quantum theory",
J. Comp. Theor. Nanosci. 8, 1011 - 1039 (2011).
http://rugth30.phys.rug.nl/pdf/DeRaedt.JCTN8.1011.2011.pdf

Marian Kupczynski

## Detailed analysis of time series

- M.K., New test of completeness of quantum mechanics, ICTP preprint IC/84/242, (1984)
- M.K., Phys.Lett. A 116(1986), 417.
- M.K., Is quantum theory predictably complete?, Phys.Scr. T135, 014005 (2009). doi:10.1088/0031-8949/2009/T135/014005
- M.K., Time series, stochastic processes and completeness of quantum theory. AIP. Conf. Proc. 1327, 394 -400 (2011)
- Box, G. E. P., Jenkins, G. M. and Reinsel, G. C.: Time Series Analysis Forecasting and Control. Wiley, Hoboken (2008)

Marian Kupczynski